\documentclass[a4paper,11pt]{article}
\pdfoutput=1

\usepackage[height=23.6truecm,width=16truecm,top=3.05truecm]{geometry}

\usepackage[skip=6pt,indent]{parskip}

\usepackage{amsmath,amssymb,amsthm,mathtools,bm,braket}
\usepackage{booktabs,array}
\usepackage{graphicx}
\usepackage{placeins}
\usepackage{float}
\usepackage{xcolor}
\usepackage{cite}
\usepackage[hyperfootnotes=false]{hyperref}
\usepackage{setspace}
\usepackage{enumitem}
\usepackage{indentfirst}
\usepackage{multirow}

\definecolor{DarkBlue}{HTML}{000080}
\hypersetup{
hypertexnames=false,
colorlinks=true,
linkcolor=DarkBlue,
urlcolor=DarkBlue,
citecolor=DarkBlue,
linktocpage,
pdftitle={Thermalization in a repeated-interaction model with memory effects},
pdfauthor={Kanji Nishii, Ryo Nagai, and Yoshihiko Abe}
}
\setlist[itemize]{topsep=0.15em,itemsep=0.05em,leftmargin=2em}
\allowdisplaybreaks
\graphicspath{{figures/}}
\numberwithin{equation}{section}

\newcommand{\Tr}{\operatorname{Tr}}

\newcommand{\Imm}{\operatorname{Im}}

\newcommand{\hit}{\mathrm{hit}}
\newcommand{\settle}{\mathrm{settle}}

\definecolor{dgreen}{rgb}{0.2,0.51,0.19}

\usepackage{ulem}

\begin{document}
\begin{titlepage}

\vspace{2.5cm}

\begin{center}

{\LARGE \bfseries
Thermalization in a repeated-interaction model
\\
with memory effects
}

\vspace{1cm}

\renewcommand{\thefootnote}{\fnsymbol{footnote}}
{%
\hypersetup{linkcolor=black}
Kanji Nishii$^{1,2}$\footnote{kanji.nishii@keio.jp},
\
Ryo Nagai$^{2,3}$\footnote{ryo.nagai.jd@hitachi.com},
and
Yoshihiko Abe$^{1,2,4}$\footnote{yabe3@keio.jp}
}%
\vspace{8mm}

{\itshape%
$^1${Graduate School of Science and Technology, Keio University, Yokohama, Kanagawa 223-8522, Japan}\\
$^2${Quantum Computing Center, Keio University, 3-14-1 Hiyoshi, Kohoku-ku, Yokohama, Kanagawa, 223-8522, Japan}\\
$^3${Research and Development Group,
Hitachi, Ltd., Kokubunji, Tokyo 185-8601, Japan}\\
$^4${Keio University Sustainable Quantum Artificial Intelligence Center (KSQAIC), Keio University, Tokyo 108-8345, Japan}\\
}

\vspace{8mm}

\end{center}

\begin{abstract}
We study how memory affects the thermalization of a qubit in a repeated-interaction model. 
In our model, a memory qubit interacts repeatedly with the system, while both qubits undergo collisions with fresh thermal ancillas. 
We derive exact closed evolution equations for their populations and mutual coherence, and demonstrate memory effects using a causal-break operation. Although the system converges to the target thermal state, the thermalization behavior shows non-monotonic approach. We also find a Mpemba-like reversal, in which the number of collisions required for thermalization depends non-monotonically on the temperature of the target state. 
Within the initial-state family considered, both phenomena require initial coherence between the system and memory or an imbalance in their excited-state populations. 
\end{abstract}

\end{titlepage}

\renewcommand{\thefootnote}{\arabic{footnote}}
\setcounter{footnote}{0}
\setcounter{page}{1}

\tableofcontents

\section{Introduction}
\label{sec:intro}

Thermal-state preparation is a key task in physics and in computer science. Thermal states are needed in a wide range of applications: thermal expectation values such as energies and correlation functions can be evaluated from thermal states prepared on a quantum simulator~\cite{2000-TerhalDiVincenzo-PRA,2011-TemmeEtAl-Nature}, and thermal states serve as the starting point of thermodynamic processes in quantum thermodynamics~\cite{2016-GooldEtAl-JPAMT,2016-VinjanampathyAnders-CP}. How to prepare thermal states efficiently on a quantum computer is therefore a question of practical importance~\cite{2023-ChenEtAl,2025-DingEtAl-CMP}, and answering it requires an understanding of the mechanism of thermalization.

\begin{figure}[t]
\centering
\includegraphics[width=0.94\linewidth]{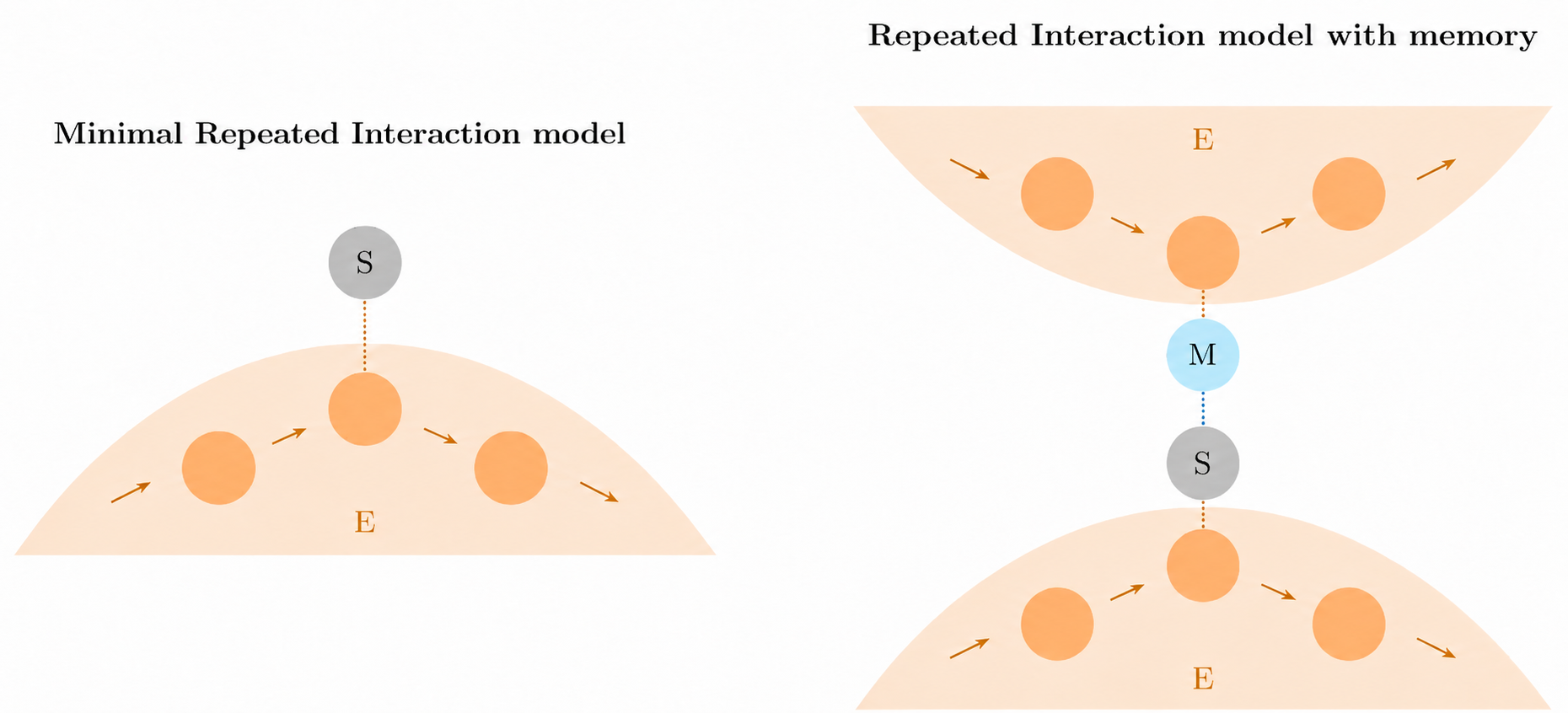}
\caption{Left panel: the minimal RI model. The system (S) collides successively with fresh ancillas (E) supplied by a thermal bath, each of which is discarded after the collision. Right panel: the extended model studied in this paper. A memory (M), which collides with S at every step, and a thermal bath of ancillas that thermalizes M are added.}
\label{fig:ri-models}
\end{figure}

The repeated-interaction (RI) model, also known as the collision model, is a simple and useful model for understanding the mechanism of thermalization~\cite{2002-ScaraniEtAl-PRL,2002-ZimanEtAl-PRA,2007-AttalJoye-JSP,2014-BruneauEtAl-JMP,2021-CampbellVacchini-E,2022-CiccarelloEtAl-PR}. In this model, the system (S) collides successively with fresh ancillas (E), which yields Markovian dynamics (Fig.~\ref{fig:ri-models}, left panel). The model has recently been used to study thermalization~\cite{2017-StrasbergEtAl-PRX,2024-BeeverEtAl-PRA,2025-ProsittoEtAl-QST,2025-Ramon-EscandellEtAl-PRR} and to design quantum algorithms, in particular for the simulation of Markovian dynamics~\cite{2021-CattaneoEtAl-PRL,2023-ErbanniEtAl-PRA,2023-PatelWilde-OSID,2023-PatelWilde-OSIDa,2024-BartolomeoEtAl-PRR,2025-HaganWiebe,2025-PocrnicEtAl-JPAMT}.

The minimal RI model (Fig.~\ref{fig:ri-models}, left panel), however, cannot describe memory effects, i.e., the dependence of the present dynamics of the system on its past history. Memory effects arise when the system interacts again with a part of the bath that retains a record of their earlier interaction, for example when an ancilla is reused or imperfectly reset. Collision models that incorporate memory have been developed to describe such non-Markovian dynamics~\cite{2009-PellegriniPetruccione-JPAMT,2012-GiovannettiPalma-PRL,2013-CiccarelloEtAl-PRA,2014-McCloskeyPaternostro-PRA,2016-KretschmerEtAl-PRA,2017-LorenzoEtAl-PRA,2018-CampbellEtAl-PRA,2024-SahaEtAl-NJP}.
In this paper, we study how memory effects affect thermalization, i.e., thermal-state preparation. We extend the minimal RI model by adding a memory (M) that retains the history of the interactions with the system (Fig.~\ref{fig:ri-models}, right panel).

In this simple model we find two nontrivial features of thermalization. First, the relaxation toward the thermal state becomes non-monotonic. Ordinarily, relaxation to a thermal state is monotonic: the system relaxes irreversibly from a hotter to a colder state, as is indeed the case in the RI model. In the extended model, in contrast, the non-Markovian dynamics (memory effects) induced by the memory allow the system to approach the target state and subsequently move away from it. Second, the number of collisions required for thermalization depends non-monotonically on the temperature of the target state; that is, a target state that is farther from the initial state (colder) can be reached in fewer collisions. This behavior is reminiscent of the Mpemba effect, in which a hotter system reaches thermal equilibrium faster than a colder one~\cite{2017-LuRaz-PNASU,2019-KlichEtAl-PRX,2021-CarolloEtAl-PRL,2024-NavaEgger-PRL,2024-WangWang-PRR,2025-ZhangEtAl-NC,2025-AresEtAl-NRP,2026-SummerEtAl-PRX}. Mpemba effects have also been reported in non-Markovian environments~\cite{2025-StrachanEtAl-PRL,2026-ZhangEtAl-PRL}, and an initial entanglement between the system and the bath itself, without any memory, has been shown to delay relaxation and give rise to a Mpemba effect~\cite{2025-Longhi-AQ}. Unlike these works, we introduce the memory into the RI model and examine how it affects the thermalization process itself.

The paper is organized as follows. In Sec.~\ref{sec:model}, we define the model, solve its dynamics exactly, and show that the model exhibits memory effects. In Sec.~\ref{sec:results}, we present the two effects of the memory on thermalization, non-monotonic thermalization and the Mpemba-like reversal, and analyze their origin. 
Sec.~\ref{sec:conclusions} is devoted to our conclusions.

\section{Repeated-interaction model with memory}
\label{sec:model}

\subsection{Setup}

\begin{figure}[t]
\centering
\includegraphics[width=0.94\linewidth]{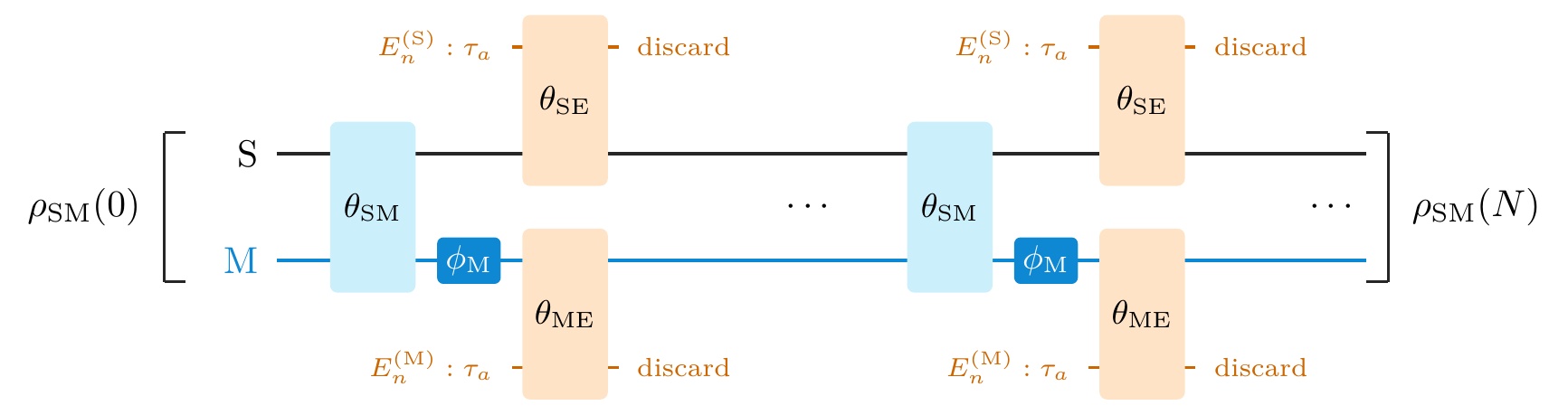}
\caption{Schematic of the model. One step consists of a partial swap between S and M (angle $\theta_{\mathrm{SM}}$), a phase rotation $R_Z^{(\mathrm{M})}(\phi_{\mathrm{M}})$ on M, a partial swap between S and a fresh ancilla $\mathrm{E}^{(\mathrm{S})}_n$ ($\theta_{\mathrm{SE}}$), and a partial swap between M and a fresh ancilla $\mathrm{E}^{(\mathrm{M})}_n$ ($\theta_{\mathrm{ME}}$); the ancillas are discarded after each step. $N$ steps map the initial state $\rho_{\mathrm{SM}}(0)$ to $\rho_{\mathrm{SM}}(N)$.}
\label{fig:our-model}
\end{figure}

In this section, we introduce our RI model with a memory, which quantum circuit is illustrated in Fig.~\ref{fig:our-model}.
For simplicity, we make the following assumptions.
The system (S) and the memory (M) are qubits. 
At each $n$-th step, two fresh ancilla qubits $\mathrm{E}^{(S)}_n$ and $\mathrm{E}^{(M)}_n$ are supplied and interact with S and M, respectively.
The aim of the ancilla introduction is to bring the system to the target thermal state
\begin{align}
\tau_a = (1-a)\ket{0}\!\bra{0} + a\ket{1}\!\bra{1}\,,
\label{eq:tau_a}
\end{align}
where $\ket{0}$ and $\ket{1}$ denote the ground and excited states.\footnote{In our minimal setup, the system is a single qubit, the same size as the ancillas, so one may object that we assume the availability of thermal qubits in order to prepare a thermal qubit. It is known, however, that a large system can be thermalized by thermal ancillas much smaller than the system~\cite{2000-TerhalDiVincenzo-PRA,2024-BartolomeoEtAl-PRR,2025-HaganWiebe}.} 
The excited-state population $a$ is determined by the inverse temperature $\beta$ and the energy gap $\Delta>0$ as
\begin{align}
a = \frac{1}{1+e^{\beta\Delta}}\,.
\label{eq:a}
\end{align}

As the initial state of our analysis, the density operator for the joint state of S and M is taken to be
\begin{align}
\rho_{\mathrm{SM}}(0) =
\begin{pmatrix}
P^{(0)}_{00} & 0 & 0 & 0 \\
0 & P^{(0)}_{01} & c_0^{*} & 0 \\
0 & c_0 & P^{(0)}_{10} & 0 \\
0 & 0 & 0 & P^{(0)}_{11}
\end{pmatrix}\,,
\label{eq:rho_SM0}
\end{align}
in the energy basis $\{\ket{00},\ket{01},\ket{10},\ket{11}\}$ of the SM system. Here, $P^{(0)}_{ij}$ are nonnegative parameters and $c_0$ is a complex parameter: the former are the populations of the energy eigenstates, and the latter is the initial coherence between $\ket{10}$ and $\ket{01}$, i.e., within the one-excited sector of S and M. 
These satisfy the following conditions:
\begin{align}
    P_{ij}^{(0)} \geq 0,
    \qquad 
    \sum_{i,j} P^{(0)}_{ij} =1,
    \qquad 
    |c_0|^2 \leq P_{01}^{(0)} P_{10}^{(0)}.
\end{align}
Note that the reduced initial state of the system obtained by tracing out the memory, $\rho_\mathrm{S}(0) = \Tr_{\mathrm{M}} \rho_{\mathrm{SM}}(0)$ differs in general from $\tau_a$. 
The task is to bring the system state close to $\tau_a$ by repeating the interactions described below.

Following the standard RI model~\cite{2002-ScaraniEtAl-PRL}, we take all two-qubit interactions to be partial-swap unitaries, which conserve the number of excitations. 
In the energy basis, the unitary operation is parametrized as
\begin{align}
U(\theta) =
\begin{pmatrix}
1 & 0 & 0 & 0 \\
0 & \cos\theta & -i\sin\theta & 0 \\
0 & -i\sin\theta & \cos\theta & 0 \\
0 & 0 & 0 & 1
\end{pmatrix}\,.
\label{eq:partial_swap}
\end{align}
We write the swap angles of the S--M, S--$\mathrm{E}^{(\mathrm{S})}_n$, and M--$\mathrm{E}^{(\mathrm{M})}_n$ interactions as $\theta_{\mathrm{SM}}$, $\theta_{\mathrm{SE}}$, and $\theta_{\mathrm{ME}}$, respectively, and take
\begin{align}
    0 < \theta_{\mathrm{SE}}, \theta_{\mathrm{ME}} < \frac{\pi}{2}\,,
\end{align}
for the interactions with the ancillas.
This condition is motivated by the convergence of the system to the target as shown in Sec.~\ref{sec:thermalization}.
In addition to the swap interactions, we apply the single-qubit phase rotation to the memory
\begin{align}
R^{(\mathrm{M})}_{Z}(\phi_{\mathrm{M}})
=
\begin{pmatrix}
e^{-i\phi_{\mathrm{M}}/2} & 0 \\
0 & e^{i\phi_{\mathrm{M}}/2}
\end{pmatrix}\,,
\label{eq:RZ}
\end{align}
immediately after the S--M partial swap.
Note that this rotation acts on the coherence stored in the memory.

As shown in Fig.~\ref{fig:our-model}, the sequence of the S--M partial swap, the phase rotation on M, and the two partial swaps with the fresh ancillas constitutes one circuit step, which corresponds to one step of the time evolution. This step is repeated $N$ times.
This operation is the CPTP map acting on $\rho_{\mathrm{SM}}$ as 
\begin{align}
\rho_{\mathrm{SM}}(N) = \Phi_a[\rho_{\mathrm{SM}}(N-1)] = \Phi_a^{N}[\rho_{\mathrm{SM}}(0)] \,,\qquad N\geq1\,,
\label{eq:Phi_a}
\end{align}
and the reduced state is given by tracing out the memory of this density operator $\rho_{\mathrm{S}}(N) = \Tr_{\mathrm{M}} \rho_{\mathrm{SM}}(N)$.

\subsection{Thermalization}
\label{sec:thermalization}

Let us now examine how the reduced state $\rho_{\mathrm{S}}(N)$ of the system evolves with $N$ and approaches the target thermal state $\tau_a$. We consider the state $\rho_{\mathrm{SM}}(N)$ of the SM system after $N$ steps and define the excited-state populations of the system and the memory,
\begin{align}
p_{\mathrm{S}}(N) = \Tr\left[(\ket{1}\!\bra{1}\otimes I)\,\rho_{\mathrm{SM}}(N)\right]\,,
\qquad 
p_{\mathrm{M}}(N) =\Tr\left[(I\otimes\ket{1}\!\bra{1})\,\rho_{\mathrm{SM}}(N)\right]\,.
\label{eq:pM}
\end{align}
When a subsystem has reached the thermal state~\eqref{eq:tau_a}, its population equals $a$. We therefore measure the deviation from the target by
\begin{align}
u_N = p_{\mathrm{S}}(N)-a\,,
\qquad v_N = p_{\mathrm{M}}(N)-a\,.
\label{eq:vN}
\end{align}
The task of the model is to achieve $u_N\simeq0$. In addition, we define the coherence between S and M after $N$ steps by
\begin{align}
c_N = \bra{10}\rho_{\mathrm{SM}}(N)\ket{01}\,.
\label{eq:cN}
\end{align}

The time evolution of \eqref{eq:Phi_a} is rewritten in terms of $(u_N, v_N, c_N)$ as
\begin{align}
\begin{pmatrix}
u_N & c_N \\
c_N^{*} & v_N
\end{pmatrix}
=
K \begin{pmatrix}
    u_{N-1} & c_{N-1} \\
    c_{N-1}^\ast & v_{N-1}
\end{pmatrix} K^\dagger
=
(K)^N
\begin{pmatrix}
u_0 & c_0 \\
c_0^{*} & v_0
\end{pmatrix}
(K^{\dagger})^N\,,
\label{eq:evolution}
\end{align}
where
\begin{align}
K =
\begin{pmatrix}
e^{-i\frac{\phi_{\mathrm{M}}}{2}}\cos\theta_{\mathrm{SE}} & 0 \\
0 & e^{i\frac{\phi_{\mathrm{M}}}{2}}\cos\theta_{\mathrm{ME}}
\end{pmatrix}
\begin{pmatrix}
\cos\theta_{\mathrm{SM}} & -i\sin\theta_{\mathrm{SM}} \\
-i\sin\theta_{\mathrm{SM}} & \cos\theta_{\mathrm{SM}}
\end{pmatrix}.
\label{eq:K}
\end{align}
Since $K$ is the product of a diagonal matrix and a unitary matrix, its operator norm is $\|K\|=\max(\cos\theta_{\mathrm{SE}},\cos\theta_{\mathrm{ME}})$, which is smaller than unity for $0<\theta_{\mathrm{SE}},\theta_{\mathrm{ME}}<\pi/2$. 
Hence $u_N$ approaches 0 at large $N$ for any initial state and any target temperature, which means that the system converges to the target thermal state $\tau_a$.
See App.~\ref{app:calculation-detail} for the details.

Eq~\eqref{eq:evolution} describes the dynamics of the system coupled to the memory.
We note that in the RI model without memory, i.e., for $\theta_{\mathrm{SM}}=0$, where the memory is decoupled from the system, Eq.~\eqref{eq:evolution} gives
\begin{align}
u_N \,=\, \cos^{2N}\theta_{\mathrm{SE}}\,u_0\,,
\label{eq:decoupled}
\end{align}
and the system relaxes monotonically to the thermal state.

\subsection{Memory effects}
\label{sec:memory}

In the previous subsection, we derive the time evolution equation of the system~\eqref{eq:evolution}.
We now show that this dynamics exhibits memory effects, i.e., that the past history of the system influences its present dynamics.

In order to characterize the memory effect, we introduce the following operation:
\begin{align}
\mathcal{C}_\sigma[\rho_{\mathrm{SM}}] = \sigma_{\mathrm{S}}\otimes \Tr_{\mathrm{S}}\rho_{\mathrm{SM}} \,,
\label{eq:def-causal-break}
\end{align}
where $\sigma_{\mathrm{S}}$ is a fixed reference state independent of the past dynamics. The operation $\mathcal{C}_\sigma$ replaces the state of the system with one that carries no information about its past. 
If the evolution of the system from $\mathcal{C}_{\sigma}[\rho_{\mathrm{SM}}]$ depends on the history before the replacement, the past information stored in the memory affects the system dynamics.
This operation is called a {\it causal break}~\cite{2018-PollockEtAl-PRL,2018-PollockEtAl-PRA}, which is also discussed in the context of non-Markovianity.

To test for memory effects, we carry out the following thought experiment. 
We consider two initial states, A and B, that differ only in the initial state of S: $\rho^{(A)}_{\mathrm{SM}}(0)=\rho^{(A)}_S(0)\otimes \rho_{\mathrm{M}}(0)$ and $\rho^{(B)}_{\mathrm{SM}}(0)=\rho^{(B)}_S(0)\otimes \rho_{\mathrm{M}}(0)$, with the same initial state $\rho_{\mathrm{M}}(0)$ of M. 
In each setup, we evolve the state for one step, obtaining $\rho^{(A)}_{\mathrm{SM}}(1)$ and $\rho^{(B)}_{\mathrm{SM}}(1)$, apply the causal break to S, and evolve for one more step. 
Immediately after the causal break, the state of S is fixed to be the same $\sigma_{\mathrm{S}}$ in both cases, so that in the absence of memory effects the populations of S after the next step would coincide. 
In our model, however, their difference is
\begin{align}
\Tr\biggl[
  (\ket{1}\!\!\bra{1}\otimes I_{\mathrm{M}})
  (\Phi_a \circ \mathcal{C}_\sigma)
  (\rho^{(A)}_{\mathrm{SM}}(1)-\rho^{(B)}_{\mathrm{SM}}(1))
\biggr]
=
\cos^2\theta_{\mathrm{SE}}
\cos^2\theta_{\mathrm{ME}}
\sin^4\theta_{\mathrm{SM}}
(u^{(A)}(0)-u^{(B)}(0))
\,,
\end{align}
which is nonzero whenever $\theta_{\mathrm{SM}}\neq0$. The difference in the initial state of S is thus stored in M and returned to S after the causal break.
This is nothing but the memory effect.

\section{Thermalization with memory effects}
\label{sec:results}

In the previous section we showed that our extended RI model (Fig.~\ref{fig:our-model}) exhibits memory effects. 
This section examines how these memory effects affect the thermalization of the system.

\subsection{Non-monotonic thermalization}
\label{sec:nonmono}

\begin{figure}[t]
\centering
\includegraphics[width=0.94\linewidth]{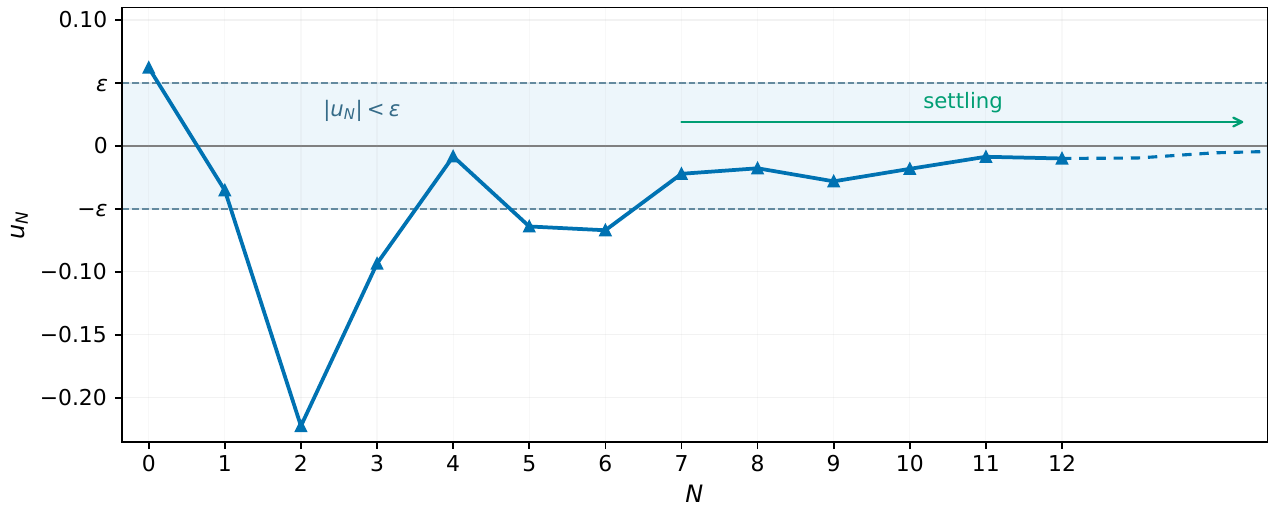}
\caption{$u_N$ as a function of the number of collisions $N$ for $\beta\Delta=0.25$, $p_{\mathrm{S}}(0)=0.5$, $p_{\mathrm{M}}(0)=0.3$, real $c_0=0.8\sqrt{0.15}$, and the representative parameter set of Sec.~\ref{sec:nonmono}. The shaded band is the region $|u_N|<\epsilon$ with $\epsilon=0.05$. The first step at which $|u_N|<\epsilon$ is $N=1$, while the first step from which $|u_N|<\epsilon$ holds at all later steps is $N=7$.}
\label{Fig02}
\end{figure}

In our model shown in Fig.~\ref{fig:our-model}, the system thermalizes through repeated collisions with the bath: as the number of collisions $N$ increases, the state of the system approaches the thermal state. 
Without the memory M, this approach is monotonic, as shown by Eq.~\eqref{eq:decoupled}.

With the memory, however, the relaxation can be non-monotonic. 
We show a typical behavior of the thermalization in Fig.~\ref{Fig02}, which plots $u_N$ as a function of $N$.
Throughout the numerical results below, we use the representative parameter set 
\begin{align}
    \Delta = 1\,,
    \qquad 
    (\theta_{\mathrm{SM}}, \phi_{\mathrm{M}}, \theta_{\mathrm{SE}}, \theta_{\mathrm{ME}}) = (\pi/4, -3\pi/10, \pi/12, \pi/4)\,,
\end{align}
and the tolerance is chosen as $\epsilon=0.05$ used to judge whether the system has reached the target (see Sec.~\ref{sec:mpemba}). 
In Fig.~\ref{Fig02}, the target and the initial condition are $\beta\Delta=0.25$, $p_{\mathrm{S}}(0)=0.5$, $p_{\mathrm{M}}(0)=0.3$, and $c_0=0.8\sqrt{0.15}$. 
Instead of decreasing monotonically to zero, $u_N$ converges to zero while oscillating: the system relaxes toward the thermal state non-monotonically. 
This behavior is specific to the model with memory.

In order to investigate the origin of the non-monotonic thermalization shown in Sec.~\ref{sec:mpemba}, let us return to the evolution equation~\eqref{eq:evolution} and find the solution of $u_N$.
In the following analysis, we take the initial state of S to be the maximally mixed state, $p_{\mathrm{S}}(0)=1/2$ (this choice is also used in the numerical results from Fig.~\ref{Fig02} on), so that $u_0=1/2-a$. 
The population deviation $u_N$ after $N$ collisions can then be parametrized as 
\begin{align}
u_N=
A_N s
+
B_N  \delta_{\mathrm{M}}
+
C_N(c_0)
\,,
\label{eq:decompose-uN}
\end{align}
with $s= 1/2 - a$ and $\delta_{\mathrm{M}} = p_{\mathrm{M}}(0) - p_{\mathrm{S}}(0)$.
For the details of the derivation, see App.~\ref{app:calculation-detail}.
The coefficients $A_N$ and $B_N$ are functions of $(\theta_{\mathrm{SM}},\theta_{\mathrm{SE}},\theta_{\mathrm{ME}},\phi_{\mathrm{M}})$, and $C_N(c_0)$ is a function of $(\theta_{\mathrm{SM}},\theta_{\mathrm{SE}},\theta_{\mathrm{ME}},\phi_{\mathrm{M}},c_0)$ with $C_N(0)=0$. 
The non-monotonic behavior of $u_N$ arises from the cancellation of the first term $A_N s$ by the second and third terms in Eq.~\eqref{eq:decompose-uN}. 
Without the memory, for example, $A_N=\cos^{2N}\theta_{\mathrm{SE}}$ and $B_N=C_N=0$, so that no cancellation occurs and $u_N$ decreases monotonically.

\begin{figure}[t]
\centering
\includegraphics[width=0.94\linewidth]{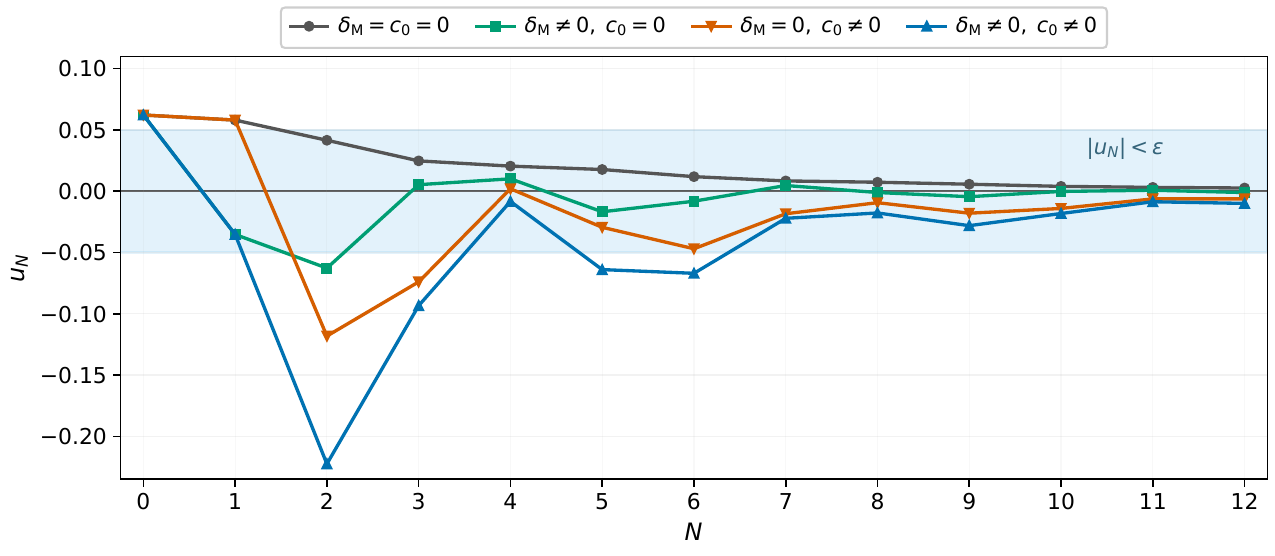}
\caption{$u_N$ for the four initial conditions listed in the text: all combinations of $p_{\mathrm{M}}(0)\in\{0.5,\,0.3\}$ ($\delta_{\mathrm{M}}=0$ or $-0.2$) and real $c_0\in\{0,\,c_\ast\}$ with $c_\ast=0.8\sqrt{0.15}$, at $\beta\Delta=0.25$, $p_{\mathrm{S}}(0)=0.5$, and the representative parameter set of Sec.~\ref{sec:nonmono}. The shaded band is as in Fig.~\ref{Fig02}. The first step at which $|u_N|<\epsilon$ and the first step from which it holds at all later steps are $(2,2)$ for $\delta_{\mathrm{M}}=c_0=0$, $(1,3)$ for $\delta_{\mathrm{M}}\neq0$ and $c_0=0$, $(4,4)$ for $\delta_{\mathrm{M}}=0$ and $c_0=c_\ast$, and $(1,7)$ when both are nonzero.}
\label{Fig03}
\end{figure}

We now confirm numerically that the second and third terms in Eq.~\eqref{eq:decompose-uN} each contribute to the non-monotonic behavior. Fig.~\ref{Fig03} shows $u_N$ for the four cases
\begin{itemize}
\item $\delta_{\mathrm{M}}=c_0=0$, black line
\item $\delta_{\mathrm{M}}\neq0\,,~~c_0=0$, green line
\item $\delta_{\mathrm{M}}=0\,,~~c_0\neq 0$, red line
\item $\delta_{\mathrm{M}}\neq0\,,~~c_0\neq 0$, blue line.
\end{itemize}
The target and $p_{\mathrm{S}}(0)$ are the same as in Fig.~\ref{Fig02}; 
for $\delta_{\mathrm{M}}\neq0$ we take $p_{\mathrm{M}}(0)=0.3$ and $\delta_{\mathrm{M}}=-0.2$, and for $c_0\neq0$ we take $c_0=0.8\sqrt{0.15}$, which is 80\% of the smaller of the two upper bounds allowed by positivity.
For $\delta_{\mathrm{M}}=c_0=0$, $u_N=A_Ns$ decreases monotonically to zero, since $A_{N+1}\leq\|K\|^2 A_N$ with $\|K\|<1$ (see App.~\ref{app:decomposition} for details). The other three curves show non-monotonic behavior for nonzero $\delta_{\mathrm{M}}$ or $c_0$.
The non-monotonic thermalization is therefore caused by $\delta_{\mathrm{M}}$ or $c_0$. We stress that adding the memory M alone does not produce non-monotonic thermalization; a nonzero $\delta_{\mathrm{M}}$ or $c_0$ is required. This observation clarifies the relation between the memory effect of Sec.~\ref{sec:memory} and the non-monotonic thermalization. The condition $\theta_{\mathrm{SM}}\neq0$ for the memory effect is also necessary for $B_N$ and $C_N(c_0)$ to be nonzero (for $\theta_{\mathrm{SM}}=0$, $B_N=C_N=0$); both conditions express the existence of a channel through which information about S returns to S via M. The existence of this channel alone, however, does not produce non-monotonic thermalization. It requires, in addition, an asymmetry in the initial state of M ($\delta_{\mathrm{M}}\neq0$ or $c_0\neq0$) that is returned to S through this channel. The same holds for the reversal of the settling order discussed in the next subsection.

We finally note that non-monotonic thermalization also occurs for other values of $\theta_{\mathrm{SM}}$ and $\phi_{\mathrm{M}}$. However, thermalization can remain monotonic even when $\delta_{\mathrm{M}}\neq0$ or $c_0\neq0$. This can occur when the decrease in $A_Ns$ at each collision is larger than any increase in $B_N\delta_{\mathrm{M}}+C_N(c_0)$ in Eq.~\eqref{eq:decompose-uN}. See App.~\ref{app:first-hit-settling} for this parameter dependence and an example of monotonic thermalization.

\subsection{Mpemba-like reversal}
\label{sec:mpemba}

\begin{figure}[t]
\centering
\includegraphics[width=0.94\linewidth]{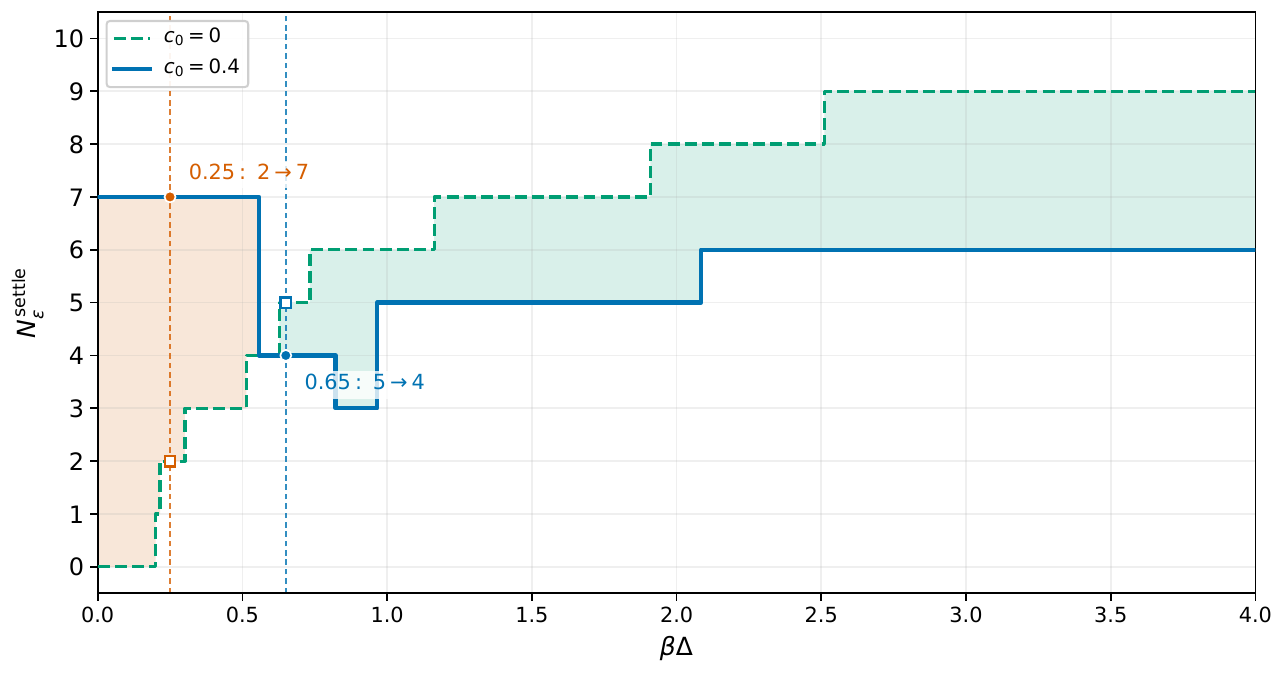}
\caption{Number of collisions $N^{\mathrm{settle}}_\epsilon$ required to reach the thermal state as a function of the inverse temperature $\beta\Delta$ of the target, with and without initial coherence, for $p_{\mathrm{S}}(0)=p_{\mathrm{M}}(0)=1/2$ ($\delta_{\mathrm{M}}=0$) and the representative parameter set of Sec.~\ref{sec:nonmono}. Dashed line: $c_0=0$; solid line: real $c_0=0.4$. The shading indicates where the solid line lies above or below the dashed line. At the marked targets $\beta\Delta=0.25$ and $0.65$, the coherence changes $N^{\mathrm{settle}}_\epsilon$ from 2 to 7 and from 5 to 4, respectively. The range shown is $0\le\beta\Delta\le4$.}
\label{Fig04}
\end{figure}

In the previous subsection we observed that the memory M can lead to non-monotonic thermalization. Here we turn to a second nontrivial phenomenon in our model.
We consider the number of collisions required to reach the thermal state as a function of the temperature $1/\beta$ of the target thermal state. 

We say that the system has reached the target state when $|u_N|<\epsilon$ for a given tolerance $\epsilon>0$. Since $\rho_{\mathrm{S}}(N)$ and $\tau_a$ are both diagonal in the energy basis, $|u_N|$ equals their trace distance. In the presence of memory, $|u_N|$ may exceed $\epsilon$ again after having fallen below it (Fig.~\ref{Fig02}). We therefore define the number of collisions required to reach the thermal state as the smallest $N$ such that $|u_n|<\epsilon$ for all $n\geq N$, 
\begin{align}
    N^{\mathrm{settle}}_\epsilon=\min\{N \in \mathbb{Z}_{\geq0} : |u_n|<\epsilon\ \text{for all}\ n\geq N\}.
    \label{eq:def-N^settle}
\end{align}
In general, this differs from the first step at which $|u_N|<\epsilon$. We use $\epsilon=0.05$ in the numerical results.

The maximally mixed initial state of S ($p_{\mathrm{S}}(0)=1/2$) corresponds to infinite temperature. The trace distance to the target, $s=1/2-a$, increases with $\beta$. Without memory, a colder target (farther from the initial state) therefore requires more collisions, so that the required number of collisions increases monotonically with $\beta$. With memory, as we now show, this is no longer the case in general.

An example is shown in Fig.~\ref{Fig04}, where the required number of collisions is plotted against the inverse temperature of the target. The two curves differ in the initial coherence $c_0$ between S and M.
The dashed line is for $c_0=0$ and the solid line for $c_0=0.4$.
The other parameters are same as those used in Sec.~\ref{sec:nonmono}.
The solid line shows that thermalization is non-monotonic with respect to $\beta\Delta$.
Moreover, a comparison of the two lines shows that $c_0$ delays thermalization at small $\beta\Delta$ (for example, $2\to7$ at $\beta\Delta=0.25$) and accelerates it at large $\beta\Delta$ ($5\to4$ at $\beta\Delta=0.65$).
This implies that the same initial coherence hinders or assists thermalization depending on the target temperature. 
Since this kind of non-monotonic behavior resembles the anomalous relaxation known as the Mpemba effect~\cite{2017-LuRaz-PNASU,2019-KlichEtAl-PRX,2021-CarolloEtAl-PRL,2024-NavaEgger-PRL,2024-WangWang-PRR,2025-ZhangEtAl-NC,2025-AresEtAl-NRP,2026-SummerEtAl-PRX}, we refer to it as a {\it Mpemba-like reversal}.

\begin{figure}[t]
\centering
\includegraphics[width=0.94\linewidth]{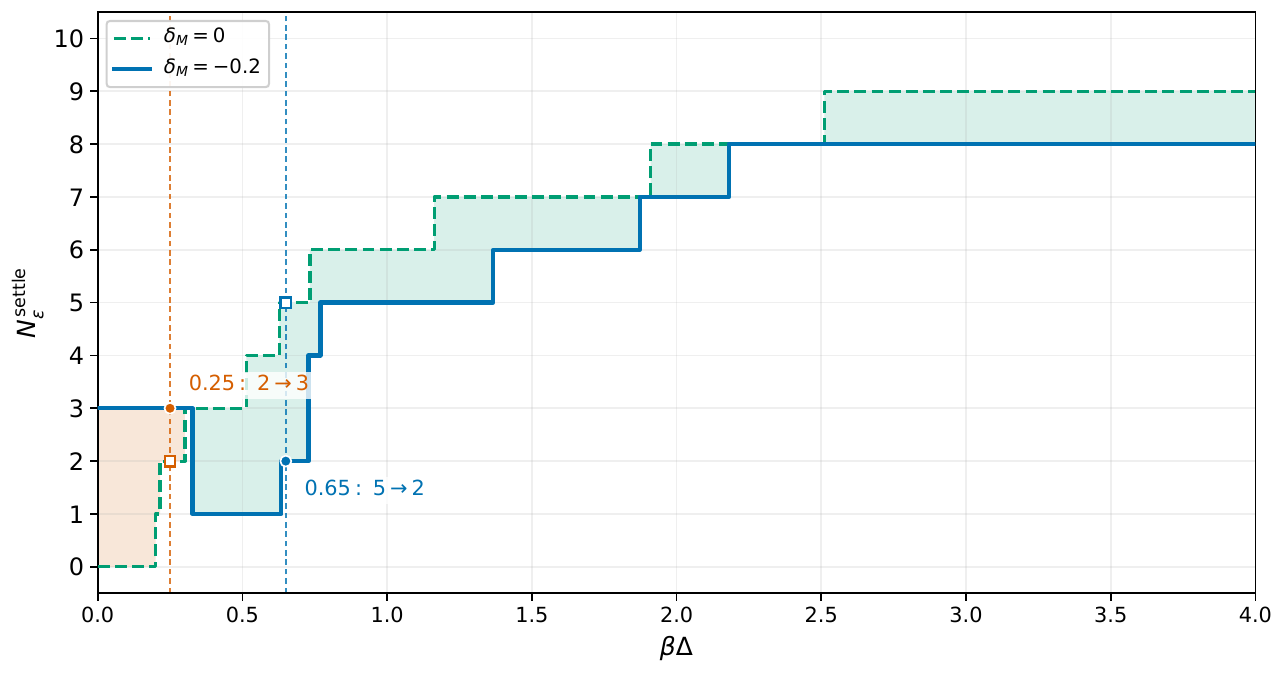}
\caption{Same as Fig.~\ref{Fig04}, but with the initial population of the memory varied instead of the initial coherence, for $p_{\mathrm{S}}(0)=0.5$, $c_0=0$, and the representative parameter set of Sec.~\ref{sec:nonmono}. Dashed line: $p_{\mathrm{M}}(0)=0.5$ ($\delta_{\mathrm{M}}=0$); solid line: $p_{\mathrm{M}}(0)=0.3$ ($\delta_{\mathrm{M}}=-0.2$). At the marked targets $\beta\Delta=0.25$ and $0.65$, $N^{\mathrm{settle}}_\epsilon$ changes from 2 to 3 and from 5 to 2, respectively. The range shown is $0\le\beta\Delta\le4$.}
\label{Fig06}
\end{figure}
We have seen that a nonzero $c_0$ produces the Mpemba-like reversal. The same phenomenon occurs for nonzero $\delta_{\mathrm{M}}$, as shown in Fig.~\ref{Fig06}, where the dashed line is for $p_{\mathrm{M}}(0)=0.5$ ($\delta_{\mathrm{M}}=0$) and the solid line for $p_{\mathrm{M}}(0)=0.3$ ($\delta_{\mathrm{M}}=-0.2$), with $p_{\mathrm{S}}(0)=0.5$, $c_0=0$, and the other parameters as in Sec.~\ref{sec:nonmono}. Here too, thermalization is delayed at $\beta\Delta=0.25$ ($2\to3$) and accelerated at $\beta\Delta=0.65$ ($5\to2$).

The Mpemba-like reversal thus also relies on the contributions of $c_0$ and $\delta_{\mathrm{M}}$, which can again be understood from the decomposition~\eqref{eq:decompose-uN} in Fig.~\ref{Fig04}. 
In the decomposition \eqref{eq:decompose-uN}, only $s=1/2-a$ in the first term depends on the target temperature, whereas the second and third terms, $B_N\delta_{\mathrm{M}}$ and $C_N(c_0)$, are the same for all targets. 
When $B_N\delta_{\mathrm{M}}+C_N(c_0)$ has the sign opposite to $A_Ns$, the two partially cancel, and the degree of cancellation depends on the target.
Tab.~\ref{tab:target-cancellation} gives examples of this cancellation for the settings in Figs.~\ref{Fig04} and~\ref{Fig06}. At the steps shown, $A_Ns$ is larger for the farther target at $\beta\Delta=0.65$, but cancellation with the remaining terms gives a smaller $|u_N|$ than for the target at $\beta\Delta=0.25$.

\begin{table}[htbp]
\centering
\caption{Contributions to $u_N$ in Eq.~\eqref{eq:decompose-uN} for the settings in Figs.~\ref{Fig04} and~\ref{Fig06}.}
\label{tab:target-cancellation}
\renewcommand{\arraystretch}{1.15}
\begin{tabular}{ccccccc}
\hline
& $(\delta_{\mathrm{M}},c_0)$
& $N$
& $\beta\Delta$
& $A_Ns$
& $B_N\delta_{\mathrm{M}}+C_N(c_0)$
& $u_N$ \\
\hline
\multirow{2}{*}{Fig.~\ref{Fig04}}
& $(0,0.4)$
& $6$ & $0.25$ & $0.012$ & $-0.076$ & $-0.064$ \\
& $(0,0.4)$
& $6$ & $0.65$ & $0.030$ & $-0.076$ & $-0.046$ \\
\hline
\multirow{2}{*}{Fig.~\ref{Fig06}}
& $(-0.2,0)$
& $2$ & $0.25$ & $0.042$ & $-0.104$ & $-0.063$ \\
& $(-0.2,0)$
& $2$ & $0.65$ & $0.105$ & $-0.104$ & $0.001$ \\
\hline
\end{tabular}
\end{table}

Finally, we discuss the relation to previous work. Ref.~\cite{2025-Ramon-EscandellEtAl-PRR} reported a similar phenomenon, in which a target farther from the initial state is reached sooner, in the RI model without memory for systems with three or more levels. There, the reversal was attributed to the dependence of the decay rate of the slowest relaxation mode on the target temperature, and it was shown that no reversal occurs for a qubit. The mechanism of the reversal in our model is different. Since the system is a qubit and the matrix $K$ in Eq.~\eqref{eq:K} does not depend on the target population $a$, the decay rates are the same for all target temperatures; the reversal originates instead from the dependence of the cancellation between $B_N\delta_{\mathrm{M}}+C_N(c_0)$ and $A_N s$ in Eq.~\eqref{eq:decompose-uN} on the target temperature. For a qubit, the memory is the only route to this reversal in our model. The role of initial system--environment correlations has been studied in Ref.~\cite{2025-Longhi-AQ}, where an initial entanglement between the system and the bath itself, without a memory, delays relaxation and produces a Mpemba effect between different initial states. In our model, the fresh ancillas cannot be correlated with S, and it is the memory M that carries the initial coherence $c_0$ with S; moreover, the comparison is made between different target temperatures for a fixed initial state.

\section{Conclusions}
\label{sec:conclusions}

In this paper we have proposed an extension of the RI model in order to analyze how memory effects, i.e., the influence of the past evolution on the present dynamics, affect thermalization. Specifically, we added a memory (M) that retains the history of the interactions, and analyzed the thermalization of a qubit analytically.

We found that the memory leads to non-monotonic relaxation of the system toward the thermal state. We also found a Mpemba-like reversal, in which the number of collisions required for thermalization depends non-monotonically on the target temperature. Importantly, neither phenomenon follows from the mere presence of the memory: both occur only in the presence of an initial coherence between the system and the memory or a difference between their initial excited-state populations.

Several directions remain for future work. While the system considered here is a single qubit, the model can be extended to many-qubit systems, and how the non-monotonic thermalization found here changes with the system size is an open question. Applications to thermal-state preparation algorithms on quantum computers are another important direction.

\section*{Acknowledgments}
\begin{sloppypar}
\noindent
The authors thank Naoki Yamamoto for helpful comments and discussions. This work was supported by MEXT Quantum Leap Flagship Program Grant No. JPMXS0120319794. K.N. also acknowledges support from MEXT Quantum Leap Flagship Program Grant No. JPMXS0118067285. Y.A. acknowledges support from JST Grant No. JPMJPF2221.
\end{sloppypar}

\appendix

\section{Evolution equation and decomposition}
\label{app:calculation-detail}

We give the detailed analysis of the evolution described by Eqs.~\eqref{eq:evolution} and \eqref{eq:K} in this appendix.

\subsection{Evolution equation}
\label{app:evolution_equation}

We begin with the derivation of the evolution equation.
Let $u_N'$, $v_N'$, and $c_N'$ denote the quantities after the S--M interaction $U(\theta_{\mathrm{SM}})$ in step $N+1$, before the phase rotation and the interactions with the thermal ancillas. Applying the partial swap in Eq.~\eqref{eq:partial_swap} gives
\begin{align}
\begin{split}
u_N'
&= \cos^2\theta_{\mathrm{SM}}\,u_N
 + \sin^2\theta_{\mathrm{SM}}\,v_N
 - \sin(2\theta_{\mathrm{SM}})\,\Imm c_N\,,
 \\
v_N'
&= \sin^2\theta_{\mathrm{SM}}\,u_N
 + \cos^2\theta_{\mathrm{SM}}\,v_N
 + \sin(2\theta_{\mathrm{SM}})\,\Imm c_N\,,
 \\
c_N'
&= \cos^2\theta_{\mathrm{SM}}\,c_N
 + \sin^2\theta_{\mathrm{SM}}\,c_N^*
 + \frac{i}{2}\sin(2\theta_{\mathrm{SM}})(u_N-v_N)\,.
\end{split}
\label{eq:app-sm-evolution}
\end{align}
The phase rotation $R_Z^{(\mathrm{M})}(\phi_{\mathrm{M}})$ in Eq.~\eqref{eq:RZ} multiplies $c_N'$ by $e^{-i\phi_{\mathrm{M}}}$. After acting $U(\theta_{\mathrm{SE}})$ and $U(\theta_{\mathrm{ME}})$, we trace out the thermal ancillas and obtain
\begin{align}
\begin{split}
u_{N+1}
&= \cos^2\theta_{\mathrm{SE}}\,u_N'\,,\\
v_{N+1}
&= \cos^2\theta_{\mathrm{ME}}\,v_N'\,,\\
c_{N+1}
&= e^{-i\phi_{\mathrm{M}}}
   \cos\theta_{\mathrm{SE}}\cos\theta_{\mathrm{ME}}\,c_N'\,.
\end{split}
\label{eq:app-thermal-evolution}
\end{align}
The thermal population $a$ does not appear explicitly because $u_N'$ and $v_N'$ are deviations from $a$. Combining Eqs.~\eqref{eq:app-sm-evolution} and~\eqref{eq:app-thermal-evolution} gives
\begin{align}
\begin{pmatrix}
u_{N+1} & c_{N+1}\\
c_{N+1}^* & v_{N+1}
\end{pmatrix}
=
K
\begin{pmatrix}
u_N & c_N\\
c_N^* & v_N
\end{pmatrix}
K^\dagger\,,
\label{eq:app-closed-evolution}
\end{align}
with $K$ given in Eq.~\eqref{eq:K}.

\subsection{Decomposition for memory effects}\label{app:decomposition}

We next derive the decomposition of Eq.~\eqref{eq:decompose-uN}. For $p_{\mathrm{S}}(0)=1/2$, the definitions $s=1/2-a$ and $\delta_{\mathrm{M}}=p_{\mathrm{M}}(0)-p_{\mathrm{S}}(0)$ give $u_0=s$ and $v_0=s+\delta_{\mathrm{M}}$, so that
\begin{align}
\begin{pmatrix}
u_0 & c_0\\
c_0^* & v_0
\end{pmatrix}
=
\begin{pmatrix}
s & c_0\\
c_0^* & s+\delta_{\mathrm{M}}
\end{pmatrix}
=
s\begin{pmatrix}
1 & 0\\
0 & 1
\end{pmatrix}
+\delta_{\mathrm{M}}
\begin{pmatrix}
0 & 0\\
0 & 1
\end{pmatrix}
+\begin{pmatrix}
0 & c_0\\
c_0^* & 0
\end{pmatrix}\,.
\label{eq:initial-response}
\end{align}
To extract the upper-left entry $u_N$ in Eq.~\eqref{eq:evolution}, we introduce the vector $e_{\mathrm{S}}=(1,0)^\top$.
Using the decomposition \eqref{eq:initial-response}, it is found that Eq.~\eqref{eq:evolution} is rewritten as
\begin{align}
u_N(s,\delta_{\mathrm{M}},c_0)
=
e_{\mathrm{S}}^\top K^N
\begin{pmatrix}
s & c_0\\
c_0^* & s+\delta_{\mathrm{M}}
\end{pmatrix}
(K^\dagger)^N e_{\mathrm{S}}
=A_Ns+B_N\delta_{\mathrm{M}}+C_N(c_0)\,,
\label{eq:affine_target_response}
\end{align}
where
\begin{align}
A_N &= e_{\mathrm{S}}^\top K^N(K^\dagger)^N e_{\mathrm{S}},
\quad
B_N = e_{\mathrm{S}}^\top K^N
\begin{pmatrix}
0 & 0\\
0 & 1
\end{pmatrix}
(K^\dagger)^N e_{\mathrm{S}},
\quad
C_N(c_0) = e_{\mathrm{S}}^\top K^N
\begin{pmatrix}
0 & c_0\\
c_0^* & 0
\end{pmatrix}
(K^\dagger)^N e_{\mathrm{S}}\,.
\label{eq:ABC_response_coefficients}
\end{align}
For $\delta_{\mathrm{M}} = c_0 =0$, $u_N$ simply becomes $A_N s$ and the expression for $A_N$ in Eq.~\eqref{eq:ABC_response_coefficients} gives
\begin{align}
A_{N+1}
= \bigl\|K^\dagger(K^\dagger)^N e_{\mathrm{S}}\bigr\|^2
\leq \|K\|^2 \bigl\|(K^\dagger)^N e_{\mathrm{S}}\bigr\|^2
= \|K\|^2 A_N\,.
\label{eq:app-monotonicity}
\end{align}
With $\|K\|<1$, this inequality shows that $u_N=A_Ns$ decreases monotonically to zero. Since $K$ is independent of $a$, $A_N$, $B_N$, and $C_N(c_0)$ do not depend on the target temperature when the interaction parameters and $c_0$ are fixed.
With $\delta_{\mathrm{M}}$ also fixed, all temperature dependence in Eq.~\eqref{eq:affine_target_response} is contained in $s=1/2-a$.

\section{Thermalization under parameter variations}
\label{app:parameters}

\subsection{Parameter dependence of non-monotonic thermalization}
\label{app:first-hit-settling}

We examine whether the non-monotonic thermalization shown in Fig.~\ref{Fig02} also occurs for other values of $\theta_{\mathrm{SM}}$ and $\phi_{\mathrm{M}}$. The settling step $N_\epsilon^{\mathrm{settle}}$ was defined by \eqref{eq:def-N^settle} in Sec.~\ref{sec:mpemba}. Here we also introduce the first hit $N_\epsilon^{\mathrm{hit}}$ and the gap $G_\epsilon$ between the two,
\begin{align}
\begin{split}
N_\epsilon^{\hit}
&= \min\{N \in \mathbb{Z}_{\geq0}: |u_N|<\epsilon\}\,,\\
N_\epsilon^{\settle}
&= \min\{N \in \mathbb{Z}_{\geq0} : |u_n|<\epsilon \text{ for all } n\geq N\}\,,\\
G_\epsilon
&= N_\epsilon^{\settle}-N_\epsilon^{\hit}\,.
\end{split}
\label{eq:app-hit-settle-gap}
\end{align}
$G_\epsilon>0$ means that thermalization is non-monotonic. Note, however, that $G_\epsilon=0$ does not necessarily mean that thermalization is monotonic.

\begin{figure}[t]
\centering
\includegraphics[width=0.45\textwidth]{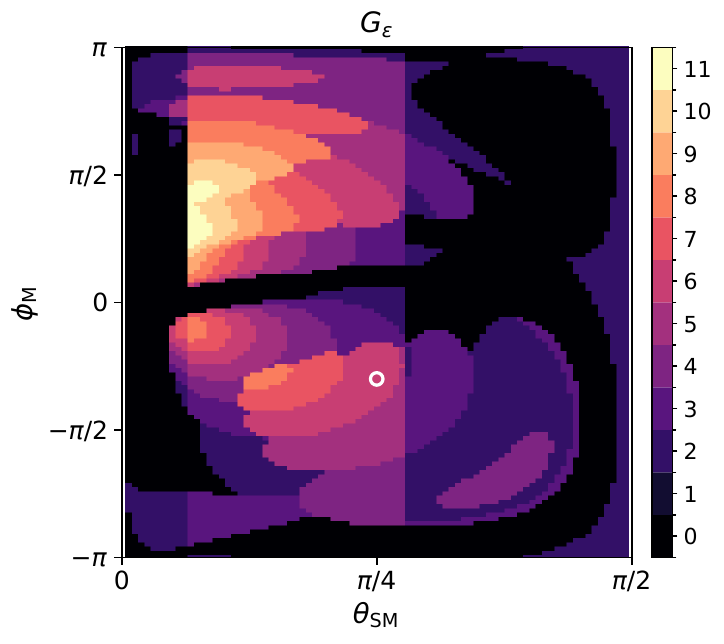}
\caption{A density plot of the gap $G_\epsilon$ as a function of $\theta_{\mathrm{SM}}$ and $\phi_{\mathrm{M}}$, with the fixed parameters given in Eq.~\eqref{eq:parameterset-app}. The white circle marks the interaction parameters used in Fig.~\ref{Fig02}.}
\label{fig:protocol-phase-diagnostics}
\end{figure}

Fig.~\ref{fig:protocol-phase-diagnostics} shows how $G_\epsilon$ depends on $\theta_{\mathrm{SM}}$ and $\phi_{\mathrm{M}}$. The initial conditions and other fixed parameters are those used in Fig.~\ref{Fig02},
\begin{align}
\begin{split}
p_{\mathrm{S}}(0)&=0.5\,,
\qquad
p_{\mathrm{M}}(0)=0.3\,,
\qquad
\delta_{\mathrm{M}}=-0.2\,,
\qquad
c_0=0.8\sqrt{0.15}\,,\\
\beta\Delta&=0.25\,,
\qquad
(\theta_{\mathrm{SE}},\theta_{\mathrm{ME}})
=(\pi/12,\pi/4)\,,
\qquad
\epsilon=0.05\,.
\end{split}
\label{eq:parameterset-app}
\end{align}
The regions with $G_\epsilon>0$ in Fig.~\ref{fig:protocol-phase-diagnostics} show that non-monotonic thermalization occurs over a broad region of the $(\theta_{\mathrm{SM}},\phi_{\mathrm{M}})$ parameter space.

Note that monotonic thermalization is also possible for the initial conditions in Eq.~\eqref{eq:parameterset-app}. For example, we set $(\theta_{\mathrm{SM}},\phi_{\mathrm{M}})=(\pi/100,0)$ while keeping the other parameters fixed to \eqref{eq:parameterset-app}. In this case, the S--M interaction is weak, and any increase in $B_N\delta_{\mathrm{M}}+C_N(c_0)$ at each collision is smaller than the decrease in $A_Ns$ in Eq.~\eqref{eq:decompose-uN}. The population deviation $u_N$ therefore decreases monotonically to zero, although both $\delta_{\mathrm{M}}$ and $c_0$ are nonzero.

\subsection{Parameter dependence of the reversal}
\label{app:mpemba-parameter-dependence}

We check whether the reversal and the acceleration or delay of thermalization shown in Sec.~\ref{sec:mpemba} also occur for other initial states and interaction parameters. We first vary the initial coherence $c_0$ at $p_{\mathrm{S}}(0)=p_{\mathrm{M}}(0)=1/2$, and then the initial populations $p_{\mathrm{S}}(0)$ and $p_{\mathrm{M}}(0)$ at $c_0=0$. In each case, we also vary $\theta_{\mathrm{SM}}$ and $\phi_{\mathrm{M}}$ while keeping the initial state fixed. Unless varied, we use $(\theta_{\mathrm{SM}},\phi_{\mathrm{M}})=(\pi/4,-3\pi/10)$, $(\theta_{\mathrm{SE}},\theta_{\mathrm{ME}})=(\pi/12,\pi/4)$, and $\epsilon=0.05$.

\subsubsection{Initial coherence}

\begin{figure}[t]
\centering
\includegraphics[width=\textwidth]{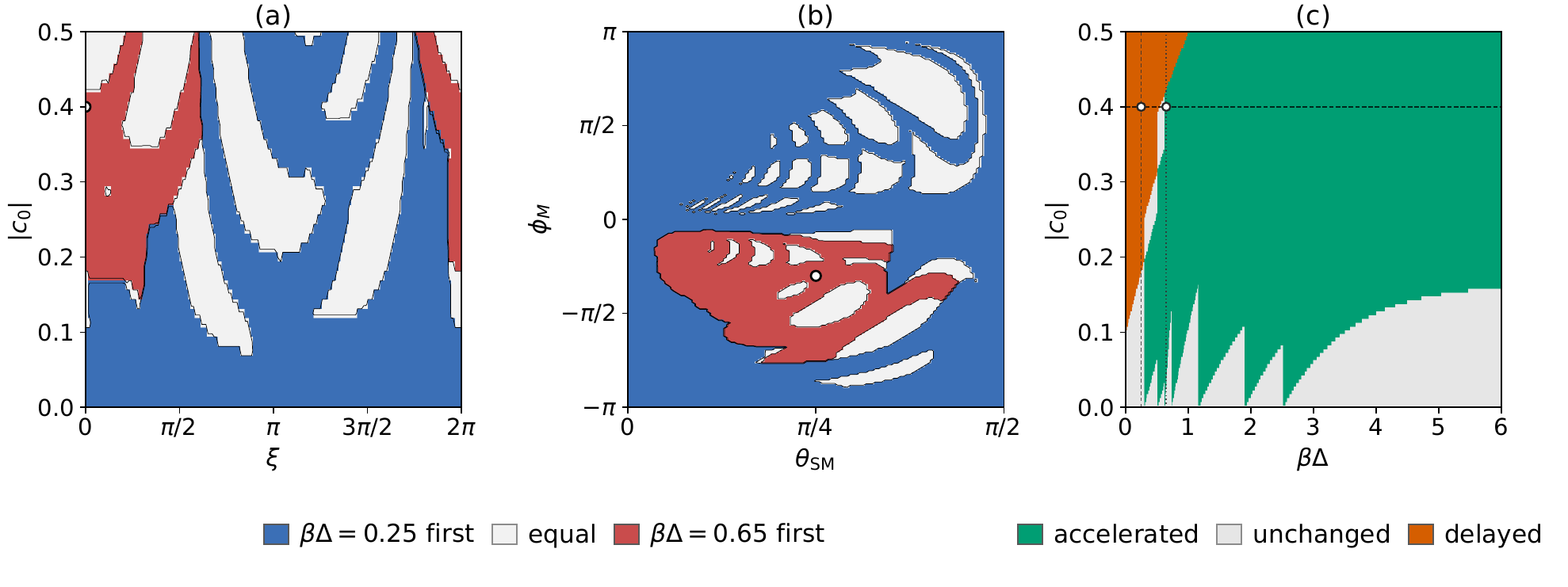}
\caption{Panels (a) and (b) show which of the two targets, $\beta\Delta=0.25$ and $0.65$, requires fewer collisions for thermalization (smaller $N_\epsilon^{\settle}$). All panels use $p_{\mathrm{S}}(0)=p_{\mathrm{M}}(0)=1/2$. Panel (a) varies the coherence phase $\xi$ and magnitude $|c_0|$, and panel (b) varies $\theta_{\mathrm{SM}}$ and $\phi_{\mathrm{M}}$ at $c_0=0.4$. Blue and red indicate that thermalization requires fewer collisions at $\beta\Delta=0.25$ and $0.65$, respectively. Light gray indicates that both targets require the same number. Panel (c) instead shows whether initial coherence accelerates or delays thermalization compared with $c_0=0$ at the same target temperature and interaction parameters. Here $(\beta\Delta,|c_0|)$ varies with $\xi=0$. Green (``accelerated''), light gray (``unchanged''), and orange (``delayed'') indicate a decrease, no change, and an increase in $N_\epsilon^{\settle}$ relative to $c_0=0$, respectively. The horizontal dashed line in (c) marks $|c_0|=0.4$, and white circles identify the settings used in Fig.~\ref{Fig04}.}
\label{fig:beta-target-robustness-app}
\end{figure}

We fix $p_{\mathrm{S}}(0)=p_{\mathrm{M}}(0)=1/2$ as in Fig.~\ref{Fig04} and vary the magnitude and phase of $c_0=|c_0|e^{i\xi}$. Positivity of the initial density matrix restricts the magnitude to $0\leq |c_0|\leq 1/2$.

Fig.~\ref{fig:beta-target-robustness-app} (a) and (b) show that reversal also occurs for values of the coherence magnitude $|c_0|$, phase $\xi$, and interaction parameters $\theta_{\mathrm{SM}}$ and $\phi_{\mathrm{M}}$ other than those used in Fig.~\ref{Fig04}. In the regions showing reversal, the farther target at $\beta\Delta=0.65$ requires fewer collisions than the target at $\beta\Delta=0.25$. Panel (c) shows that coherence magnitudes other than $|c_0|=0.4$ can also accelerate thermalization at one target temperature and delay it at another.

\subsubsection{Initial populations}
\label{subsec:population-grid-robustness}

\begin{figure}[t]
\centering
\includegraphics[width=\textwidth]{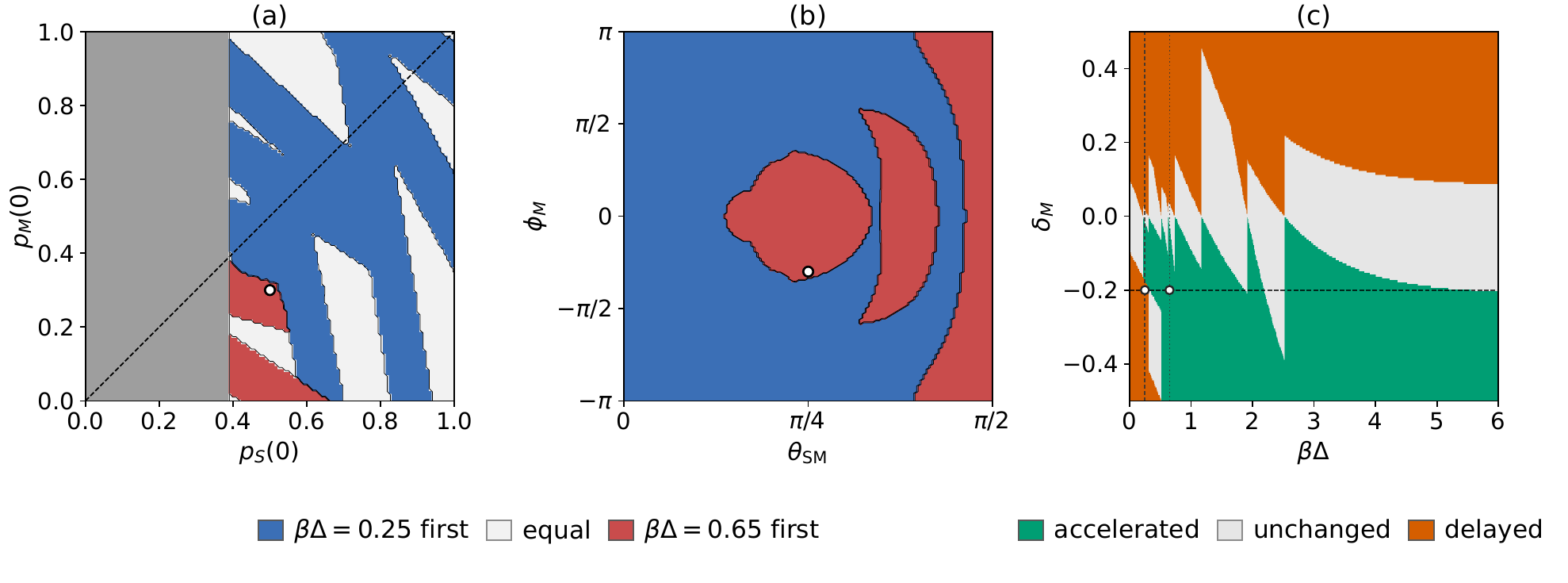}
\caption{Panels (a) and (b) show which of the two targets, $\beta\Delta=0.25$ and $0.65$, requires fewer collisions for thermalization (smaller $N_\epsilon^{\settle}$). All panels use $c_0=0$. Panel (a) varies the initial populations $p_{\mathrm{S}}(0)$ and $p_{\mathrm{M}}(0)$, and panel (b) varies $\theta_{\mathrm{SM}}$ and $\phi_{\mathrm{M}}$ at $p_{\mathrm{S}}(0)=0.5$ and $p_{\mathrm{M}}(0)=0.3$. Blue and red indicate that thermalization requires fewer collisions at $\beta\Delta=0.25$ and $0.65$, respectively. Light gray indicates that both targets require the same number. The gray region in (a) is excluded because it does not satisfy the initial-distance condition in Eq.~\eqref{eq:app-population-ordering}. Panel (c) instead shows whether an initial population imbalance $\delta_{\mathrm{M}}$ accelerates or delays thermalization compared with $p_{\mathrm{M}}(0)=0.5$ at the same target temperature and interaction parameters. Here $(\beta\Delta,\delta_{\mathrm{M}})$ varies with $p_{\mathrm{S}}(0)=0.5$ and $p_{\mathrm{M}}(0)=0.5+\delta_{\mathrm{M}}$. Green (``accelerated''), light gray (``unchanged''), and orange (``delayed'') indicate a decrease, no change, and an increase in $N_\epsilon^{\settle}$ relative to $p_{\mathrm{M}}(0)=0.5$, respectively. The horizontal dashed line in (c) marks $\delta_{\mathrm{M}}=-0.2$, and the vertical guides mark the two target temperatures. White circles identify the settings used in Fig.~\ref{Fig06}.}
\label{fig:population-seeded-robustness}
\end{figure}

With $c_0=0$, we vary the initial populations $p_{\mathrm{S}}(0)$ and $p_{\mathrm{M}}(0)$ and the interaction parameters $\theta_{\mathrm{SM}}$ and $\phi_{\mathrm{M}}$. 
When we make a plot in $(p_{\mathrm{S}}(0),p_{\mathrm{M}}(0))$-plane, we restrict $p_{\mathrm{S}}(0)$ so that the target at $\beta\Delta=0.65$ remains farther from the initial system state than the target at $\beta\Delta=0.25$, as in Fig.~\ref{Fig06}. This requires
\begin{align}
|p_{\mathrm{S}}(0)-a_\mathrm{f}|
&>
|p_{\mathrm{S}}(0)-a_\mathrm{c}|
\quad\Longleftrightarrow\quad
p_{\mathrm{S}}(0)>\frac{a_\mathrm{c}+a_\mathrm{f}}{2}\,,
\label{eq:app-population-ordering}
\end{align}
where $a_\mathrm{c}$ ($\beta\Delta=0.25$, closer) and $a_\mathrm{f}$ ($\beta\Delta=0.65$, farther) are the excited-state populations of the two targets, given by Eq.~\eqref{eq:a}.

Fig.~\ref{fig:population-seeded-robustness}(a) and (b) show that reversal also occurs for values of the initial populations $p_{\mathrm{S}}(0)$ and $p_{\mathrm{M}}(0)$ and the interaction parameters $\theta_{\mathrm{SM}}$ and $\phi_{\mathrm{M}}$ other than those used in Fig.~\ref{Fig06}. Panel (c) shows that initial population imbalances other than $\delta_{\mathrm{M}}=-0.2$ can also accelerate thermalization at one target temperature and delay it at another.

\bibliographystyle{yautphysm}
\bibliography{refs}

\end{document}